\documentclass[journal=jacsat,manuscript=article]{achemso}

\usepackage[colorlinks=true,bookmarks=false,citecolor=blue,urlcolor=blue]{hyperref} 
\setkeys{acs}{maxauthors=0}
\usepackage[version=3]{mhchem} 
\usepackage{graphicx}
\graphicspath{ {./images/} }
\usepackage{color}
\usepackage{xcolor}

\usepackage{natbib}
\usepackage[normalem]{ulem}
\usepackage{caption}
\DeclareCaptionLabelFormat{adja-page}{#1 #2}
\usepackage{ulem}
\usepackage{subfig}

\title{Photoinduced transition from quasi-2D Ruddlesden-Popper to 3D halide perovskites for optical writing multicolor and light-erasable images}

\author{Sergey S. Anoshkin}
\affiliation[ITMO University]
{ITMO University, Kronverkskiy pr. 49, 197101 St. Petersburg, Russia}
\author{Ivan I. Shishkin}
\affiliation[ITMO University]
{ITMO University, Kronverkskiy pr. 49, 197101 St. Petersburg, Russia}
\author{Daria I. Markina}
\affiliation[ITMO University]
{ITMO University, Kronverkskiy pr. 49, 197101 St. Petersburg, Russia}
\author{Lev S. Logunov}
\affiliation[ITMO University]
{ITMO University, Kronverkskiy pr. 49, 197101 St. Petersburg, Russia}
\author{Hilmi Volkan Demir}
\affiliation[Bilkent University]
{UNAM-Institute of Materials Science and Nanotechnology, National Nanotechnology Research Center, Department of Electrical and Electronics Engineering, Department of Physics, Bilkent University, Ankara, 06800, Turkey}
\alsoaffiliation[NTU]
{School of Materials Science and Engineering, Nanyang Technological University, Singapore 639798}
\author{Andrey L. Rogach}
\affiliation[City
University of Hong Kong]
{Department of Materials Science and Engineering and Centre for Functional Photonics (CFP), City
University of Hong Kong, 83 Tat Chee Avenue, Kowloon, Hong Kong SAR 999077, P. R. China}
\author{Anatoly P. Pushkarev}
\email{anatoly.pushkarev@metalab.ifmo.ru}
\affiliation[ITMO University]
{ITMO University, Kronverkskiy pr. 49, 197101 St. Petersburg, Russia}
\author{Sergey V. Makarov}
\email{s.makarov@metalab.ifmo.ru}
\affiliation[ITMO University]
{ITMO University, Kronverkskiy pr. 49, 197101 St. Petersburg, Russia}
\alsoaffiliation{Qingdao Innovation and Development Center, Harbin Engineering University, Qingdao 266000, Shandong, P. R. China}

\begin{document}
\newpage
\begin{abstract}
Development of advanced optical data storage, information encryption, and security labeling technologies requires low-cost materials exhibiting local, pronounced, and diverse modification of their structure-dependent optical properties under external excitation. Herein, for these purposes, we propose and develop a novel platform relying on layered lead halide Ruddlesden-Popper (quasi-2D) phases that undergo a light-induced transition towards bulk (3D) halide perovskite and employ this phenomenon for the direct optical writing of various multicolor patterns. This transition causes the weakening of quantum confinement, and hence the bandgap reduction in these photoluminescent thin films. To significantly extend the color gamut of evolving photoluminescence, we make use of mixed-halide compositions exhibiting photoinduced halide segregation. As a result, the emission wavelength of the resulting films can be widely tuned across the entire 450$-$600~nm range depending on the illumination conditions. We show that pulsed near-infrared femtosecond laser irradiation provides high-resolution direct writing, whereas continuous-wave ultraviolet exposure is suitable for fast recording on larger scales. The luminescent micro- and macro-scale images created on such quasi-2D perovskite films can be erased during the visualization process, by which the persistence of these images to UV light exposure can be controlled and increased further with the increasing number of octahedral layers used in the perovskite stacks. This makes the proposed writing/erasing perovskite-based platform suitable for the manufacturing of both inexpensive optical data storage devices and light-erasable security labels.                     
\end{abstract}


\section{Introduction}
As powerful tools for the manipulation of big data, state-of-the-art technologies enabling the recording of information at nano- and micro-scales are expected to speed up progress in natural sciences and medicine.~\cite{tsai2021calibration,lemm2021machine,plata2021achieving,lee2020statistical,jiang2022big}. Also, it is profoundly beneficial if such a new technology not only affords huge capacity for data storage but also enables devices with high-level security anti-counterfeit labels. From this point of view, optical data writing applied to luminescent materials such as lead halide perovskites (LHPs) could offer solutions~\cite{huang2020reversible,zhizhchenko2020light,zhan2021situ,sun2022three}. The writing process should, however, be high-throughput and well-controlled, but at the same time hard to reproduce, without having full knowledge of the specific process~\cite{cheng2020self, gong2022enzyme, larin2021luminescent}. 

Bearing all these points in mind, the choice of laser writing on solution-processed LHP thin films could be a viable option to address the required demands. Indeed, extremely low thermal conductivity ($\sim$0.5~W/m$\cdot$K, which is less than that of silica glasses~\cite{haeger2020thermal}) of ABX$_3$ (A = MA$^+$ - methylammonium, FA$^+$ - formamidinium, Cs$^+$; B = Pb$^{2+}$; X = Cl$^-$, Br$^-$, I$^-$) perovskites allows for high-density optical data writing~\cite{zhan2021situ}, as well as directly patterning laser microdisks, nanowire arrays ~\cite{zhizhchenko2019single,zhizhchenko2020light} and various micro-optical elements~\cite{zhizhchenko2021direct,zhizhchenko2021directional,wang2020flat} obtained by laser ablation. Moreover, photoluminescence (PL) spectral tunability over the visible spectral range can be achieved for LHPs via partial or complete substitution of a certain anion (e.g., Br$^-$) for another one (e.g., Cl$^-$ or I$^-$) at the X site~\cite{protesescu2015nanocrystals,liashenko2019electronic}. Moreover, both photoinduced compositional and structural changes of the perovskite lattice could significantly impact its optical properties as well. The former has been previously demonstrated for mixed-halide perovskites revealing a phenomenon of light-driven reversible anion segregation expressed in PL redshift upon illumination and its recovery in the dark~\cite{hoke2015reversible,brennan2017light,knight2018electronic}, whereas the latter is particularly true for quasi-2D Ruddlesden-Popper (RP) phase undergoing dramatic changes in structural and optical properties under intense photoexcitation~\cite{ha2019toward,leung2020mixed,sen2022uv}.

Importantly, lead halide RP phase possessing A$'_2$A$_{n-1}$Pb$_n$X$_{3n+1}$ structure (A$'$ $-$ large cation, e.g., BA$^+$ $-$ n-butylammonium, PEA$^+$ $-$ phenethylammonium) provides the possibility of tuning the PL spectrum with an additional degree of freedom owing to the quantum confinement phenomenon in thin [PbX$_6$]$^{4-}$ octahedral layers separated by A$'$ species in the perovskite stack~\cite{gan2021photophysics}, whose thickness (i.e., number of the octahedral layers $n$) is regulated by the ratio between the corresponding halide salts employed for reaction. Concerning the data storage applications, (PEA)$_2$MA$_{n-1}$Pb$_n$I$_{3n+1}$ phases with $n = 1-5$ were utilized for recording of information in resistive random access memory (ReRAM) devices~\cite{solanki2019interfacial}. Besides, it was reported that the manipulation of
light irradiation and gate voltage allows one to control the ion migration process and the current flow in a data storage device based on ambipolar SnO transistor with (PEA)$_2$PbI$_4$ photoactive layer~\cite{tian2021flexible}, and, thus, to perform writing and erasing processes under illumination only. However, the use of lead halide RP phases, offering unique optical properties related to broadband spectral tunability, for all-optical data storage and labeling has not been reported to date.

In this work, we systematically study the local optical response of mixed-halide RP phases BA$_{2}$MA$_{n-1}$Pb$_{n}$(Br,I)$_{3n+1}$ ($n = 1-3$) to light irradiation under ambient conditions, and employ its tunability for the direct optical writing of various multicolor patterns with high resolution and broadband gamut. Quasi-2D structure of spin-coated perovskite thin films is confirmed by powder X-ray diffraction (XRD) and steady-state absorption spectroscopy. It is established that linear photoexcitation invokes almost no temporal change in the PL spectrum of $n=1$ film, whereas $n=2$ and $n=3$ ones exhibit irreversible emission tunability in the 450-600 nm range. The latter is explained by light-induced compositional and structural modification of the quasi-2D phases, which experience the halide segregation, then the destruction of iodine-rich phases accompanied with I$_2$ release, and thereafter undergo a phase transition toward bulk MAPbBr$_3$ perovskite. The proposed mechanism for this observed optical response is in good agreement with $in~situ$ XRD measurements and tests demonstrating the I$_2$ release. The larger the number of octahedral layers in the perovskite stack, the more resistant the RP phase is to the aforementioned light-driven modifications. Taking all of these findings into account, as proof-of-concept demonstrations, we employ the complex behavior of evolving PL under optical irradiation in $n=3$ film for two-photon direct laser writing (DLW) of high-resolution multicolor microimages, whereas simple projection UV lithography applied to $n=2$ film yielded luminescent labels, which can be erased by light during the reading process. The developed writing/erasing perovskite-based platform leverages on cost-efficient synthesis and fabrication, and can be employed for the manufacturing of optical data storage devices, information encryption, and security labeling.

\section{Results and discussion}

\textbf{\textit{Photoinduced phase transition in Ruddlesden-Popper perovskite films.}}
Quasi-2D mixed-halide perovskite thin films with compositions of BA$_2$PbBr$_2$I$_2$ ($n=1$), BA$_2$MAPb$_2$Br$_3$I$_4$ ($n=2$), and 
BA$_2$MA$_2$Pb$_3$Br$_4$I$_6$ ($n=3$) are obtained from 0.3 M solutions of PbI$_2$, methylammonium bromide (MABr) and/or n-butylammonium bromide (BABr) mixtures in anhydrous dimethyl sulfoxide (DMSO) spin-casted onto glass substrates, followed by annealing on a hot plate (for details, see $Methods$). Quasi-2D structure of the deposited films is confirmed by XRD measured in Bragg-Brentano geometry (Fig.~\ref{Fig_1}a). Comparison of the recorded patterns with ones reported by Stoumpos et al.~\cite{stoumpos2016ruddlesden} allows us to clearly identify the number ($n$) of [PbX$_6$]$^{4-}$ octahedral layers in the perovskite-like stacks. In the 4-20$^o$ 2$\theta$ range, the first sample shows three peaks among which the most intensive one at 6.35$^o$ corresponds to the scattering of X-ray from (002) crystallographic planes confining a single octahedral layer and double layer of BA spacer cation. According to the Bragg's law, the distance between the (002) crystallographic planes ($d$-spacing) equals 13.9 {\AA}, which is slightly shorter than that in BA$_2$PbI$_4$ ($d_{(002)}$=14.2 {\AA})~\cite{stoumpos2016ruddlesden} because the mixed-halide counterpart possesses a more compact crystal lattice. Introducing one and two more octahedral layers along with a required amount of MA species into the perovskite unit results in the shifting of three diffraction peaks towards small angles, as well as the appearance of additional peaks in the XRD patterns assigned to (080) and (0100) planes, respectively (Fig.~\ref{Fig_1}a). 

Since our study is focused on optical encryption, it is important to examine the structural evolution of any sample upon UV excitation. Therefore, $in~situ$ XRD measurements for the BA$_2$MA$_2$Pb$_3$Br$_4$I$_6$ film illuminated with continuous-wave UV light ($\lambda$ = 365 ~nm, I = 72 ~mW cm$^{-2}$) for 240 min are conducted. A set of patterns collected with 40 min time interval reveals the simultaneous descending of peaks belonging to the quasi-2D phase and their shifting towards large angles (Fig.~\ref{Fig_1}b). Here it is found that the structural properties of the illuminated films do not recover in the dark conditions. Possible reasons for such irreversible behavior could be the sintering of perovskite units, divided by large cation spacer, and the loss of iodide ions promoting the formation of exclusively Br-rich structural species having smaller lattice constants as compared to that of the initial mixed-halide perovskite. Finally, all the peaks belonging to the quasi-2D phase completely disappear and the XRD pattern appears very similar to that of the bulk ($n$~=~$\infty$) cubic MAPbBr$_3$ exhibiting signals at 2$\theta$ of ca. 15$^o$ and 30$^o$, which correspond to (100) and (200) planes, respectively (Fig.~\ref{Fig_1}b). A complete change in the XRD patterns is complemented by optical images of the sample emitting in red at the beginning of the experiment and green at the end (Fig. S1).

The absorption spectrum of BA$_2$PbBr$_2$I$_2$ shows a strong exciton peak at 449 nm. With the increasing number of octahedral layers ($n$), this peak undergoes a bathochromic shift and becomes gradually less pronounced (Fig.~\ref{Fig_1}c). The former effect is caused by the band gap narrowing due to the reduction of quantum confinement in $n$~=~2 and $n$~=~3 layered RP phases, along with a slight increase in the I:Br ratio ($R$) taking the following values: 1 for $n$~=~1, $\sim$1.33 for $n$~=~2, and 1.50 for $n$~=~3. The latter effect of the weakening of exciton oscillator strength is related to the reduction of exciton binding energy with an increase in both $n$ and $R$ as well. To establish some key energetic properties of the studied materials we do deconvolution of the absorption spectra by using a function describing excitonic absorption~\cite{liashenko2019electronic} and, then, draw Tauc plots for band-to-band absorption (for details, see $Supporting~ Information$, Fig. S1). This gives us the following values for spectral broadening ($\Gamma$) and exciton binding energy ($E_b$), and band gap energy ($E_g$): for $n$~= 1, $\Gamma$ = 0.11 eV, $E_b$ = 0.22 eV, $E_g$ = 2.98 eV; for $n$~= 2 -- 0.07 eV, 0.17 eV, and 2.62 eV, respectively; for $n$~= 3 -- 0.06 eV, 0.15 eV, 2.45 eV, respectively. The descending trends for all the values with an increase in $n$ could be explained as follows: i) increase in the number of [PbX$_6$]$^{4-}$ octahedral layers gives more rigid structures that are more resistant to deformation of the crystal lattice and, hence, have a reduced number of defect states scattering excitons \cite{cortecchia2017broadband}; ii) as the thickness of 2D structure increases, the crystal lattice field more efficiently screens photoexcited electron and hole from each other and, therefore, exciton binding energy is getting lower \cite{chernikov2014exciton}; iii) the reduction of quantum confinement in $n$~= 2 and $n$~= 3 layered phases along with the increase in the I:Br ratio results in band gap narrowing. Concerning the latter, band gap energy value for BA$_2$MA$_2$Pb$_3$Br$_4$I$_6$ ($n$~= 3) is found to be much larger than that of bulk MAPbBrI$_2$ ($n$-$\rightarrow$ $\infty$) which is close to 1.8 eV~\cite{cui2016color}.

\begin{figure}[t!]
  \includegraphics[width=\linewidth]{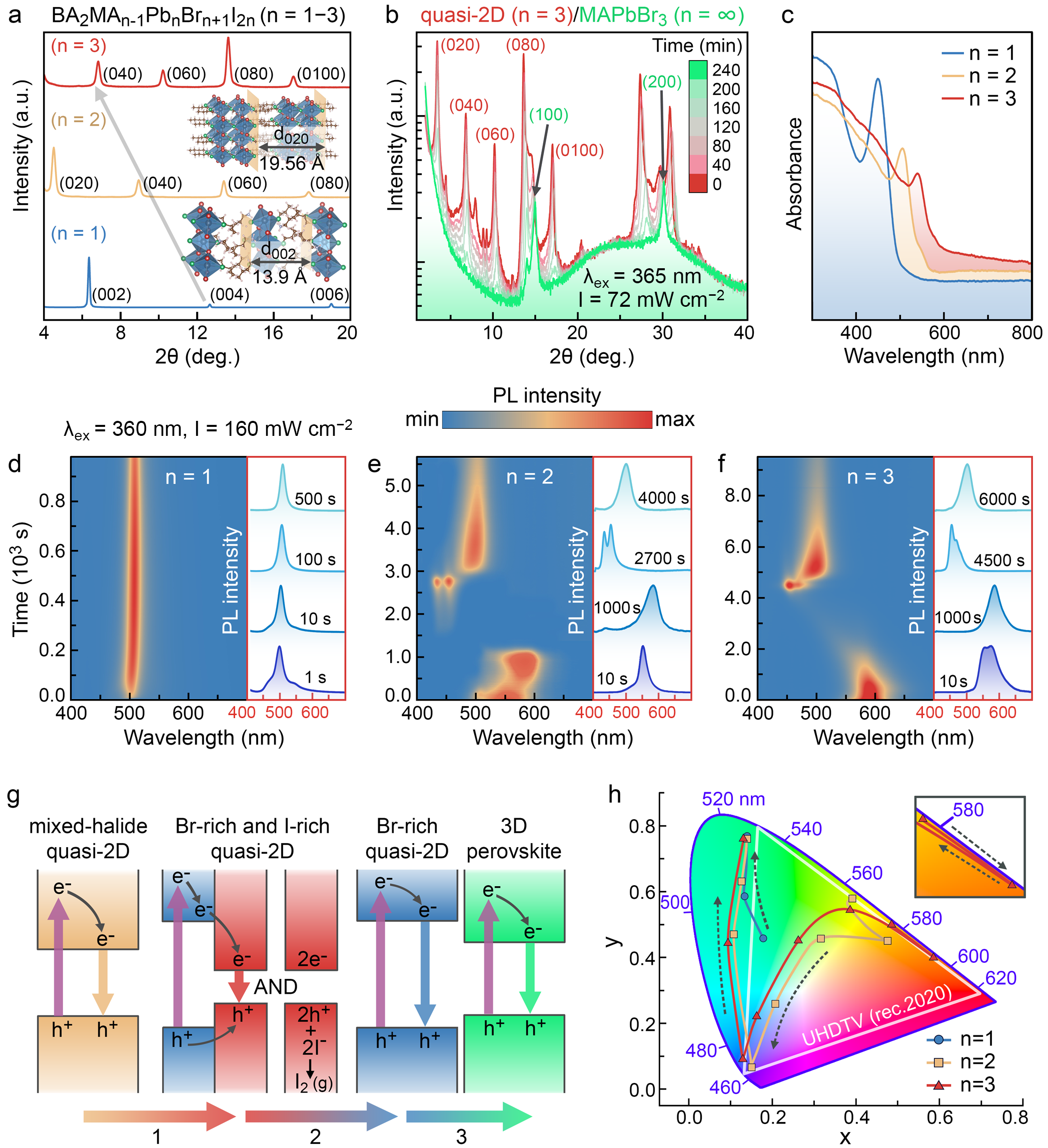}  \caption{(a) XRD patterns of BA$_2$MA$_{n-1}$Pb$_n$Br$_{n+1}$I$_{2n}$ ($n$ = 1$-$3) thin films. Insets illustrate crystal structure and similar crystallographic planes for $n$ = 1 and $n$ = 2 quasi-2D phases. The gray arrow shows an almost linear shift of the diffraction peaks towards smaller angles with increase in the number ($n$) of [PbX$_6$]$^{4-}$ octahedral layers. (b) Data derived from $in situ$ XRD measurements. According to the series of patterns, $n$ = 3 film undergoes phase transition from the quasi-2D structure to bulk MAPbBr$_3$ ($n$ = $\infty$) upon continious-wave UV excitation (I  = 72 mW cm$^{-2}$) for 240 min. (c) Absorption spectra for the $n$ = 1$-$3 RP films.}
  \end{figure}
\begin{figure}
  \captionsetup{labelformat=empty}
  \ContinuedFloat
  \caption{(d-f) Evolution of PL spectra in the films with $n = 1-3$ exposed to 360 nm light with intensity of 160 mW cm$^{-2}$. The $n$ = 1 film shows exclusively green emission from the grain boundaries, whereas $n$ = 2 and $n$ = 3 films demonstrate remarkable spectral tuning of PL color. (g) Energy diagram depicting the possible mechanism for spectral tuning in the $n$ = 2 and $n$ = 3 films. Three types of compositional and structural transformation are indicated with gradient arrows: 1) light-induced halide segregation invoking Br- and I-rich quasi-2D phases; 2) I$_2$ release-assisted dissociation of the I-rich phase and 3) sintering of Br-rich quasi-2D phase resulting in the formation of 3D bromide perovskite. (h) CIE 1931 color space exhibiting the change of PL color corresponding to the dynamics plotted in (d)-(f).}
  
  \label{Fig_1}
\end{figure}

Next PL spectral evolution for the samples is systematically studied upon illumination under 360 nm UV lamp with an intensity of 160 mW cm$^{-2}$. The films with $n$~=~1 exhibit single peak green emission peaking at 500 nm that slowly shifts towards 510 nm over the whole time interval of exposure to UV light (Fig.~\ref{Fig_1}d). Interestingly, here the spectral positions of this emission and the exciton absorption peak (Fig.~\ref{Fig_1}c) are very different and the Stokes shift, which is supposed to be small in quasi-2D perovskites~\cite{deng2020long}, is unusually large in this case. We assume that a reasonable explanation for this observation could be an energy transfer process. In particular, we reckon that excitons formed in the phase with $n$~=~1 transfer to the one with the highest $n$ formed at grain boundaries and experience radiative recombination~\cite{qin2020spontaneous,qin2022dangling}. Thus, traces of the low band gap high-$n$ phase possibly act as excitonic funnels. Taking this into account, it is not surprising that neither the XRD patterns nor the absorption spectra demonstrate the presence of the high-$n$ phase in the BA$_2$PbBr$_2$I$_2$ film. The films with $n$~=~2 and $n$~=~3 reveal similar temporal behavior of the PL signal. Initially, they emit orange light at ca. 560 nm undergoing a gradual redshift almost up to 600 nm for 1,000 s (Fig.~\ref{Fig_1}e,f), with halide segregation phenomenon being responsible for such a clearly observable change~\cite{hoke2015reversible}. The segregation leads to the formation of Br-rich and I-rich quasi-2D nanoscopic domains in perovskite films. The latter have a lower band gap than the former ones, and thus, takes an active part in the exciton funneling process. It should be pointed out that PL from the I-rich phase slowly deteriorates with time. This happens because of the dissociation of the I-rich domains invoked by the reaction 2I$^-$ + 2h$^+$ $\longrightarrow$ I$_2$ (g) describing the generation of a volatile iodine species via the oxidation of iodide ions by photoexcited holes (h$^+$) in the valence band of perovskites~\cite{brennan2020superlattices}. To find evidence of such a dissociation mechanism, the $n$~=~3 BA$_2$MA$_2$Pb$_3$Br$_4$I$_6$ film encapsulated together with a small piece of a KI test paper is illuminated until the complete deterioration of the reddish-orange luminescence (Fig. S3). One can see that the released I$_2$ gas reacts with KI, giving a darkened edge of the test paper beside the area of the film exposed to UV light for 100 min (bottom image in Fig. S3). When the I-rich domains are finally depleted, exciton radiative recombination occurs within the Br-rich ones. Therefore, for $n$~=~2 film, we observe a peak at 434 nm assigned to BA$_2$MAPb$_2$Br$_7$~\cite{leung2020mixed} accompanied with the second peak at 454 nm, most likely related to the emission from hollow perovskite structures~\cite{worku2020hollow} containing large BA cations sealed in MAPbBr$_3$ crystal lattice. These hollow perovskites are an intermediate of the sintering process, which are formed as a result of the evolution of $n$~=~3 film, along with the second less pronounced emission peaked at 468 nm belonging to BA$_2$MA$_2$Pb$_3$Br$_{10}$ phase~\cite{qin2022dangling}. Finally, blue emission in both $n$~=~2 and $n$~=~3 films changes to green, which occurs within the Br-rich quasi-2D domains due to the sintering process and results in the formation of phases with larger $n$ and finally bulk MAPbBr$_3$ ($n$~=~$\infty$) in the end. To illustrate the above discussion, a qualitative energy diagram (Fig.~\ref{Fig_1}g) visualizes three key stages during the evolution of PL color governed by the aforementioned UV light-induced structural transformations $n$~=~2 and $n$~=~3 perovskite films.

To find more evidence of the proposed mechanism we demonstrate synchronous absorption-PL dynamics for all the films (Fig. S4). First, we measure the initial absorption spectra for them and, then, they are illuminated by UV light ($\lambda_{ex}$ = 360 nm, I = 160 mW cm$^{-2}$) for obtaining PL dynamics. By interrupting the exposure of the films to UV light at certain moments and measuring absorption spectra for them consistent data on absorption and PL dynamics are collected. It is established that $n$ = 1 film exhibits the following changes in its absorption spectrum (Fig. S4a): i) the absorption peak at 449 nm goes down and undergoes a blueshift; ii) a shoulder appears in the 500-520 nm range and further evolves into a pronounced peak at 510 nm; iii) a new peak at ca. 400 nm arises. The first two observations are in good agreement with the halide segregation phenomenon, dissociation of the I-rich phase due to the release of I$_2$ species, and the formation of bulk perovskite, whereas the third one can be explained by the formation of BA$_2$PbBr$_4$ phase exhibiting its own excitonic absorption~\cite{li2020unusual}. Concerning the PL dynamics for $n$ = 1 (Fig. S4b), we emphasize it is not capable of giving clear information about structural transformations because long-range exciton transport in quasi-2D perovskites~\cite{deng2020long} and efficient energy funneling phenomena~\cite{lei2020efficient} results in photoluminescence of the bulk phase impurity evenly distributed in the film. On the contrary, absorption and PL dynamics for $n$ = 2 and $n$ = 3 look quite consistent. In particular, after the halide segregation and dissociation of the I-rich phase, in both absorption and PL spectra the peaks assigned to BA$_2$MAPb$_2$Br$_7$ ($\lambda_{abs}$ = 430 nm and $\lambda_{em}$ = 434 nm for $n$ = 2 film in Fig. S4c,d)~\cite{li2020unusual} and, most likely, to hollow perovskite structures ($\lambda_{abs}$ = 450 nm and $\lambda_{em}$ =454 nm for $n$ = 2 and $n$ = 3 films in Fig. S4c-f)~\cite{worku2020hollow} can be clearly identified.

 The evolution of PL spectra for the studied thin films can be well captured and conveniently represented in the form of shifting emission color coordinates according to CIE 1931 color space diagram (Fig.~\ref{Fig_1}h). While the sample with $n=1$ shows no significant color change, PL color for $n=2$ and $n=3$ films start from yellow (x = 0.395, y = 0.557) and orange (x = 0.559, y = 0.435), moving to purple (x = 0.167, y = 0.110) and blue (x = 0.131, y = 0.096), and, finally, turn to the same green color (x = 0.141, y = 0.775), respectively, on the color gamut. It should be noted, that both of these phases experience segregation-driven color change expressed in the saturated orange hues (x = 0.441, y = 0.476) for $n=2$ and (x = 0.591, y = 0.402) for $n=3$ at the initial stage of excitation. It is important that the $n=3$ phase displays the widest color span which extends beyond the rec. 2020 (UHDTV) color space and, hence, is preferable for the production of optical data storage devices as well as for the optical writing of anti-counterfeit images. The $n=2$ phase undergoes faster spectral evolution as compared to the former one and could provide optical images with an additional degree of security related to the limited number of reading cycles. In the following, we will demonstrate the realization of the both opportunities.

\begin{figure}[t!]
  \includegraphics[width=\linewidth]{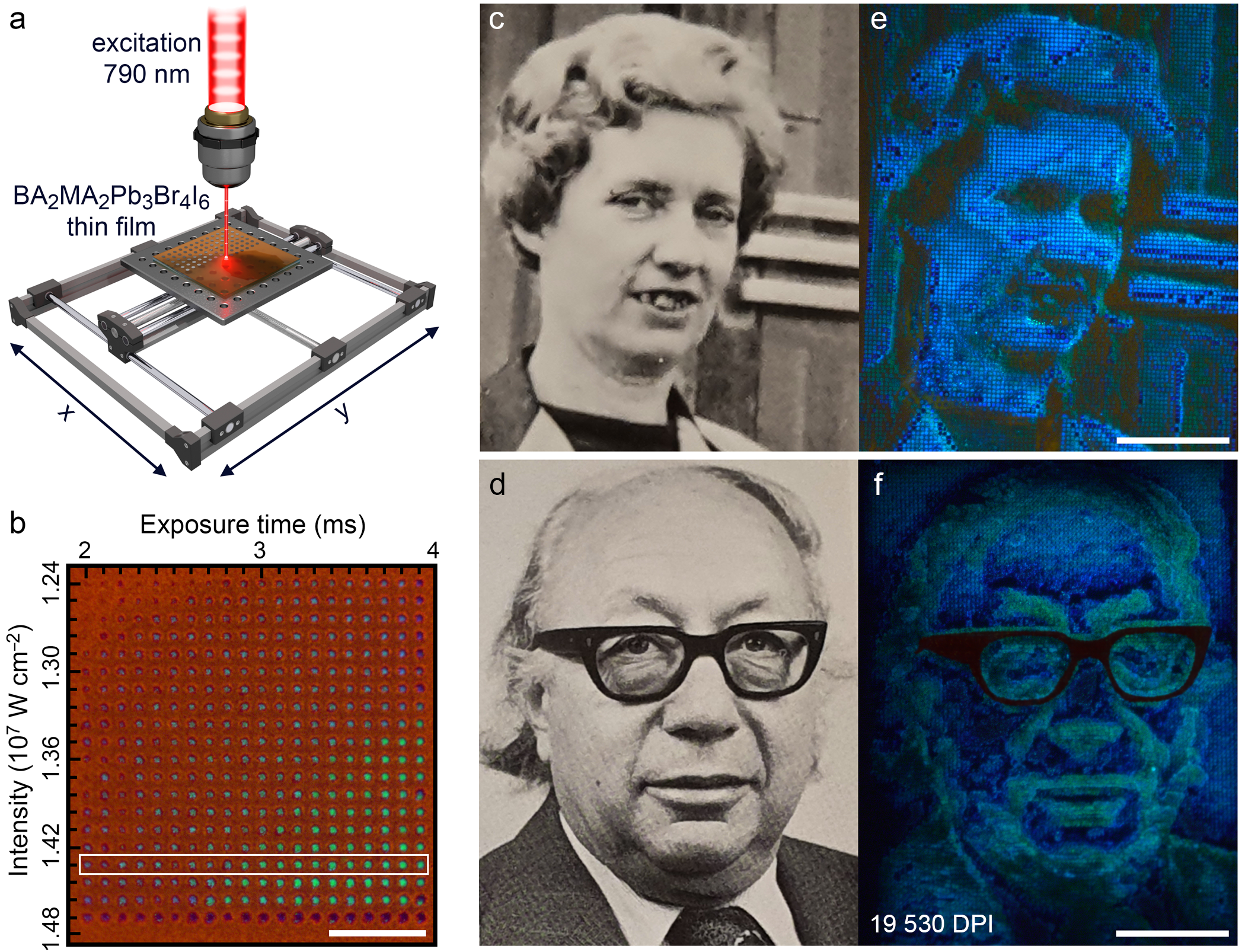}
  \caption{(a) Schematic illustration of two-photon direct laser writing (DLW) procedure applied to $n=3$ quasi-2D perovskite thin film. (b) Luminescent image of an array of microspots preliminary exposed to various intensities of 790-nm pulsed laser irradiation for various time intervals (scale bar is 50~$\mu$m). A white box identifies an optimal dynamic range for DLW giving the widest PL color gamut with no burnt (overexposed) or poorly resolved (underexposed) spots. (c-f) Original pictures of Dr. S. Ruddlesden (c) and Prof. P. Popper (d) chosen for the production of corresponding high-resolution (19500 DPI) multicolor luminescent micro-images (e,f) by using DLW method (scale bar is 50~$\mu$m).}
  \label{Fig_2}
\end{figure}

\textbf{\textit{Direct laser writing of multicolor micro-images.}}
Experimental data on the evolution of PL color for the $n=3$ quasi-2D perovskite films upon prolonged UV-light illumination (see Fig.~\ref{Fig_1}h) inspires and justifies their applicability to the production of multicolor PL patterns through the spatially resolved automatic management of irradiation dose (fluence). One of the state-of-the-art techniques for the creation of such patterns is direct laser writing (DLW)~\cite{zhan2021situ}, schematically illustrated in Figure~\ref{Fig_2}a. To implement DLW on the $n=3$ phase, we employ a lithography setup (for details, see $Methods$) based on a femtosecond pulsed Ti:sapphire laser ($\lambda_{ex}=$790 nm, $\tau=$100 fs, $f=$80 MHz). Using this DLW system, the aforementioned photoinduced structural transformation from the quasi-2D to 3D perovskite phase occurs because of the two-photon absorption phenomenon in the near-infrared range~\cite{walters2015two}. The control over the irradiation dose can conveniently be exercised by varying two parameters - intensity of the optical excitation and duration of the film exposure to laser pulses. The dynamic range (DR) for the dosage is established as follows: at an intensity varying from 1.24$\times10^7$ up to 1.47$\times10^7$ W cm$^{-2}$, the time interval of exposure varies in the range of 2.0$-$3.9 ms and subsequently the optimal set of produced luminescent pixels is selected (Fig.~\ref{Fig_2}b). Inspection of a whole array of the pixels depicted in Figure~\ref{Fig_2}b reveals that the optimal DR spans from 2.86$\times$10$^4$ to 5.75$\times$10$^4$ J cm$^{-2}$, as this range does not result in any burnt (overexposed) or poorly resolved (underexposed) pixels.

Thereafter, we conduct the dots per inch (DPI) test to evaluate the maximum capacity of information produced by DLW procedure. A bright-field micro-image (Fig. S5a) visualizing a set of 2$\times$2 dot patterns reveals that the dots with a diameter of 1 $\mu$m can be spatially resolved when the distance between their centers is equal to 1.6 $\mu$m. This provides us with a maximum resolution of ca. 19500 DPI a resolution at which we show the construction of a simple quick response (QR) code consisting of pixels emitting the same blue light (Fig. S5b,c). The maximum resolution is also further used for the creation of complex multicolor micro-images.

To demonstrate the applicability of the $n=3$ perovskite film to the optical data storage and anti-counterfeit labeling, two black and white pictures of the discoverers of the quasi-2D phase Dr. Sheila Ruddlesden and Prof. Paul Popper (Fig.~\ref{Fig_2}c,d) are reproduced on a micro-scale. For this purpose, their photos are digitized and then downsampled to 20 shades of gray. Pixels possessing the lightest shade specify the regions to be subjected to the highest irradiation dose for parts of the film resulting in a green color of PL, whereas the ones with the darkest shade indicated the unexposed area exhibiting a reddish-orange PL. The rest of the shades set up a linear increment of the fluence in the 0$-$5.75$\times$10$^4$ J cm$^{-2}$ range. As a result, corresponding multicolor high-resolution fluorescence images are formed (Fig.~\ref{Fig_2}d,f). These images provide reasonable examples of high-capacity optical data storage, since each of the 20 shades encodes the corresponding PL spectrum, which could be assigned to a certain binary number. Furthermore, the spatial distribution of the resultant luminescent pixels makes it possible to consider such images as micro-scale anti-counterfeit labels.

\begin{figure}[t!]
 \centering
  \includegraphics[width=1.0\linewidth]{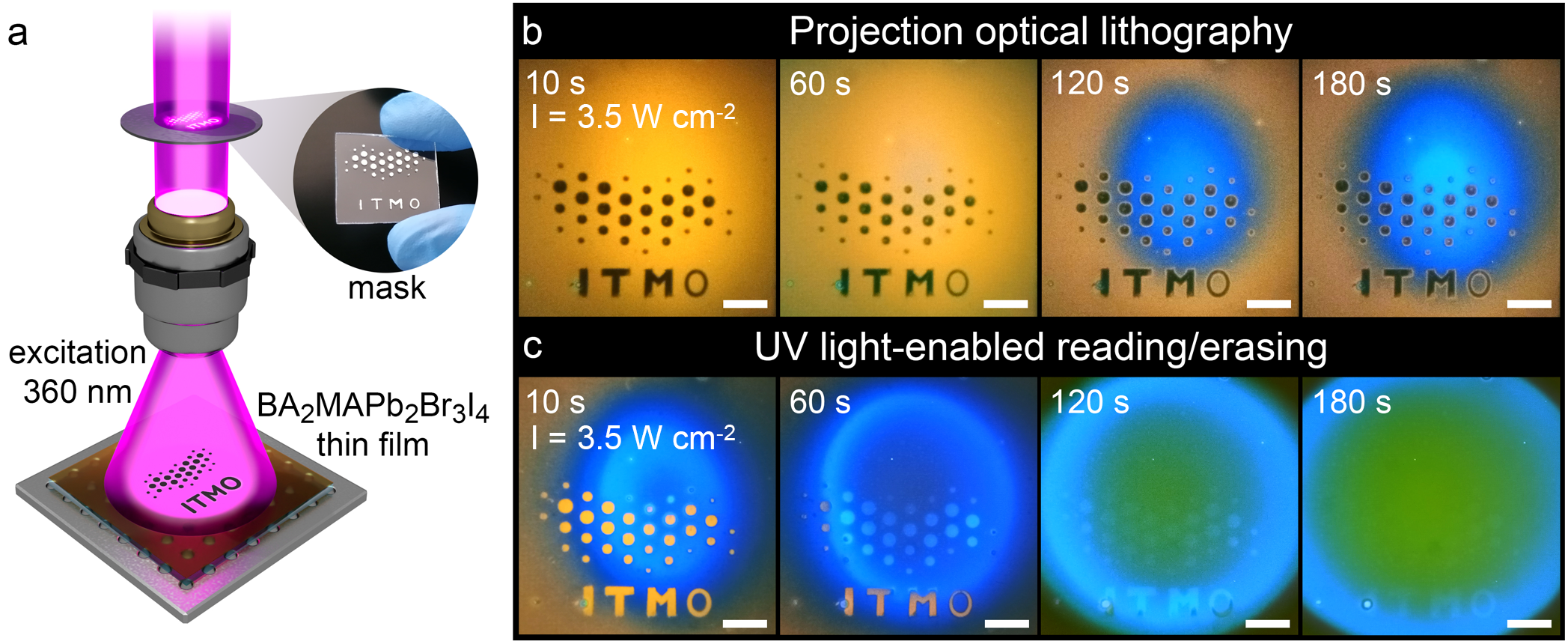}
  \caption{(a) Schematic illustration of the projection optical lithography procedure applied to the $n=2$ quasi-2D perovskite thin film. (b) Luminescent images of the area exposed through a mask (with a logotype of ITMO University) to focused UV light with intensity of 3.5 W cm$^{-2}$ for 10, 60, 120, and 180 s (scale bar is 50~$\mu$m). Black areas are shaded with the mask. Note that a blue luminescent spot appears in the center of the area because of uneven distribution of the incident light intensity over the field of view. (c) Luminescent images of the same area after the mask removal upon the same photoexcitation for 10, 60, 120, and 180 s. In the beginning of visualization process, a contrast yellow pattern within the blue spot is revealed, which gradually disappears over time and, finally, turns to green.}
  \label{Fig_3}
\end{figure}

\textbf{\textit{Projection optical lithography of light-erasable micro-images.}}
Following the trend of technological diversification for manufacturing luminescent micro-labels, we examine a simple projection optical lithography (POL) instead of DLW. For this purpose, a mask with a logotype of ITMO University consisting of a patterned aluminum film on a glass substrate is mounted before a 50$\times$ objective and shades a continuous-wave beam of UV light at 360 nm projected on the top of the $n=2$ quasi-2D film (Fig.~\ref{Fig_3}a). At the incident light intensity of 3.5 W cm$^{-2}$, the illuminated area shown in Figure~\ref{Fig_3}b becomes blue in the center after 120 s because of the uneven lateral distribution of the irradiation. After the mask removal, unexposed regions exhibit yellow luminescence and turn blue as well after 60 s of illumination (Fig.~\ref{Fig_3}c). Remarkably, more prolonged exposure to the UV light (longer than 120 s) causes erasing of the label manifested in the appearance of a green spot undergoing the temporal lateral propagation (Fig.~\ref{Fig_3}c). The control over time intervals for writing and erasing steps can be realized by tuning both the excitation intensity and thickness of the studied quasi-2D film. Such behavior of the $n=2$ phase is favorable for advanced security labeling \cite{yu2020recent,zhuang2020organic} which limits the number of label examination acts to only a few or even just one, making it invalid for further reading.

\textbf{\textit{Long-term stability of luminescent patterns.}} 
Finally, we study the resistance of encapsulated micro-images and films to photobleaching caused by long-lasting light exposure to evaluate their applicability for storage on the shelf. For this purpose, a micro-image encapsulated at ambient conditions (relative humidity 35$\%$) by using a double-sided tape and coverslip of 150 $\mu$m thickness is aged on a windowsill for 1 month. Thus, it is exposed to indoor lighting and direct sunlight in an ad hoc fashion at temperature varying in the 20--40~$^o$C range. A comparison of pictures taken of the as-prepared micro-image and the aged one shows no significant change (Fig. S6a,b). Furthermore, we measure absorption and PL spectral dynamics for encapsulated at the same conditions $n$ = 3 thin film illuminated by continuous wave UV light ($\lambda_{ex}$ = 365~nm, I = 80~mW cm$^{-2}$) for 600 min (Fig. S6c,d). It is established that the encapsulated sample demonstrates the slowing down of the spectral dynamics by more than 40 times as compared to that of non-encapsulated one irradiated by two times more intensive light (Fig. S4e,f). Importantly, spectral dynamics for $n$ = 3 film encapsulated with KI test paper is about 15 faster than that of one without it (Fig. S7). We assume the reason for such a drastic difference in dynamics stems from the ability of released I$_2$ species to react with the film and, hence, to slow down the conversion of mixed-halide quasi-2D phase into pure bromide quasi-2D one. Therefore, in the presence of KI test paper absorbing I$_2$ the conversion goes faster. Thus, I$_2$ release process is recognized to be responsible for the most pronounced change in PL color of our samples. A further change in PL color caused by the light-induced transition from pure bromide quasi-2D phase to bulk perovskite could be oxygen- or/and moisture-assisted as well as the thermally-assisted process. Considering all this, one can see how even a very simple encapsulation procedure substantially increases the temporal stability of the luminescent micro-images and films and provides them with an opportunity for commercialization.

\section{Conclusions}
In summary, we have demonstrated the light-induced PL emission tuning of BA$_2$MA$_{n-1}$Pb$_n$(Br,I)$_{3n+1}$ (n = 2, 3) thin films. The mechanisms for such spectral tuning in the studied mixed-halide quasi-2D Ruddlesden-Popper perovskites involve irreversible compositional and structural modifications, which go beyond the conventional halide segregation in 3D (bulk) perovskites. It has been established that the $n=3$ film is more resistant to light-driven transition from the quasi-2D phase to the 3D bulk phase, as compared to $n=2$ film. This has allowed us to create high-resolution multicolor micro-images on the $n=3$ quasi-2D perovskite thin films using a two-photon direct laser writing method by controlling the irradiation dose and exposure time. These images can be considered as examples of high-capacity optical data storage, and could meet the demands of information encryption technology, which significantly overcomes previous works on halide perovskites in terms of DPI~\cite{zou2019nonvolatile,minh2021low,kanwat2022reversible}. Moreover, a simple projection UV lithography has been applied to the $n=2$ quasi-2D perovskite films to demonstrate light-erasable luminescent labels, which may provide secure items with an advanced level of protection and speed up the potential progress in anti-counterfeiting.

\section{Methods}
\textbf{\textit{Fabrication of quasi-2D perovskite thin films}}. Lead halide RP phase precursor solutions were obtained by mixing 138.3 mg (0.3 mmol) of lead(II) iodide (PbI$_2$, TCI, 99.99\%, trace metals basis) with the corresponding amounts of methylammonium bromide (MABr, Greatcell Solar, 99.99\%) and/or n-butylammonium bromide (BABr, Greatcell Solar, 98\%), and dissolution of the mixtures in 1 mL of anhydrous dimethyl sulfoxide (DMSO,
Sigma-Aldrich, 99.9\%). For BA$_2$PbBr$_2$I$_2$ ($n=1$) solution, 92.4 mg (0.6 mmol) of BABr was used. For BA$_2$MAPb$_2$Br$_3$I$_4$ ($n=2$) solution, 46.2 mg (0.3 mmol) of BABr and 16.8 mg (0.15 mmol) of MABr were employed. For BA$_2$MA$_2$Pb$_3$Br$_4$I$_6$ ($n=3$) solution, 30.8 mg (0.2 mmol) of BABr and 22.4 mg (0.2 mmol) of MABr were used. Glass substrates were cleaned by subsequent ultrasonication in acetone and 2-propanol, rinsed with deionized water and exposed to ozone treatment for 10 min in order to improve wettability of the surface. The precursor solutions were spin-casted on the substrates at 2,500 rpm for 5 min and then annealed on a hot plate at 60~$^o$C for 5 min. The thickness of the resultant thin films was ~60 nm. All the procedures were conducted inside a N$_2$-filled glove box with both O$_2$ and H$_2$O concentrations (not exceeding 1 ppm).  

\textbf{\textit{Characterization of thin films}}. Thickness of the deposited films was measured by using a Tencor P-7 stylus profilometer (KLA). Optical absorption spectra were recorded using a UV-3600 spectrophotometer (Shimadzu). XRD patterns of the samples were measured in $\theta-\theta$ geometry on a
SmartLab diffractometer (Rigaku) equipped with a 9 kW rotating Cu
anode X-ray tube. Bright-field and fluorescence images of the samples were obtained on
Axio Imager A2m (Carl Zeiss) microscope with 50$\times$ and
100$\times$ objectives (Carl Zeiss EC Epiplan-NEOFLUAR). Evolution of PL spectra was recorded by using a QE Pro optical fiber
spectrometer (Ocean Optics) coupled with the microscope in
the fluorescence mode.

\textbf{\textit{Direct laser writing}}. For creating various micro-patterns on the perovskite film surface, direct laser writing (DLW) technique was applied. Laser pulses of duration 100~fs from a Ti-sapphire oscillator emitting at a wavelength of 790~nm with a repetition rate of 80~MHz were focused through a 40$\times$ objective (NA=0.7) onto the quasi-2D perovskite film surface. The irradiation power was monitored by thermopile sensor and controlled by a motorized half-wave plate ($\lambda$/2) and a polarizing beamsplitter. The exposure time was controlled by acousto-optical modulator used as a fast laser shutter. The positioning of the sample was performed by air bearing linear motor stages (Aerotech). DLW was conducted at ambient conditions (relative humidity 35$\%$).

\textbf{\textit{Projection UV optical lithography}}. A thin-film (100 nm) metal pattern on a glass substrate was fabricated by thermal evaporation of aluminum pellets (Ted Pella, Inc, 99.999\%) with a deposition rate of 0.3 nm$\cdot$s$^{-1}$ in a vacuum chamber (Kurt J Lesker Company) at
2$\times$10$^{-7}$ Torr pressure. The obtained mask with a logotype of ITMO University was mounted on the microscope instead of an aperture-adjustable optical iris diaphragm in front of HBO 100 W/2 mercury short-arc lamp (OSRAM). Optical power of the focused incident UV light was measured with a Star Bright power meter (Ophir Photonics). Projection UV optical lithography was conducted at ambient conditions (relative humidity 35$\%$).

\begin{acknowledgement}
This research was supported by Priority 2030 Federal Academic Leadership Program and by the Ministry of Science and Higher Education of the Russian Federation (Project 075-15-2021-589), and by the Croucher Foundation of Hong Kong SAR. HVD gratefully acknowledges support from TUBA. The authors thank Dr. Vidas Pak\v{s}tas for measuring XRD patterns (Fig. 1b). The authors are grateful to Ann Pace (Lucideon Ltd.) for providing them with pictures of Dr. S. Ruddlesden and Prof. P. Popper. The authors thank Mr. Ivan Pustovit for assistance in graphic design.

\end{acknowledgement}

\begin{suppinfo}
Photographs of $n=3$ quasi-2D perovskite film at the beginning and end of $in~situ$ XRD measurement, pictures of $n=3$ film examined for I$_2$ gas release, bright-field and luminescence micro-images of 2$\times$2 dots patterns and high-resolution QR code produced by using two-photon direct laser writing method.

\end{suppinfo}

\bibliography{achemso-demo}

\end{document}







\renewcommand{\figurename}{Figure S}

\begin{figure*}
\includegraphics[width=1.0\columnwidth]{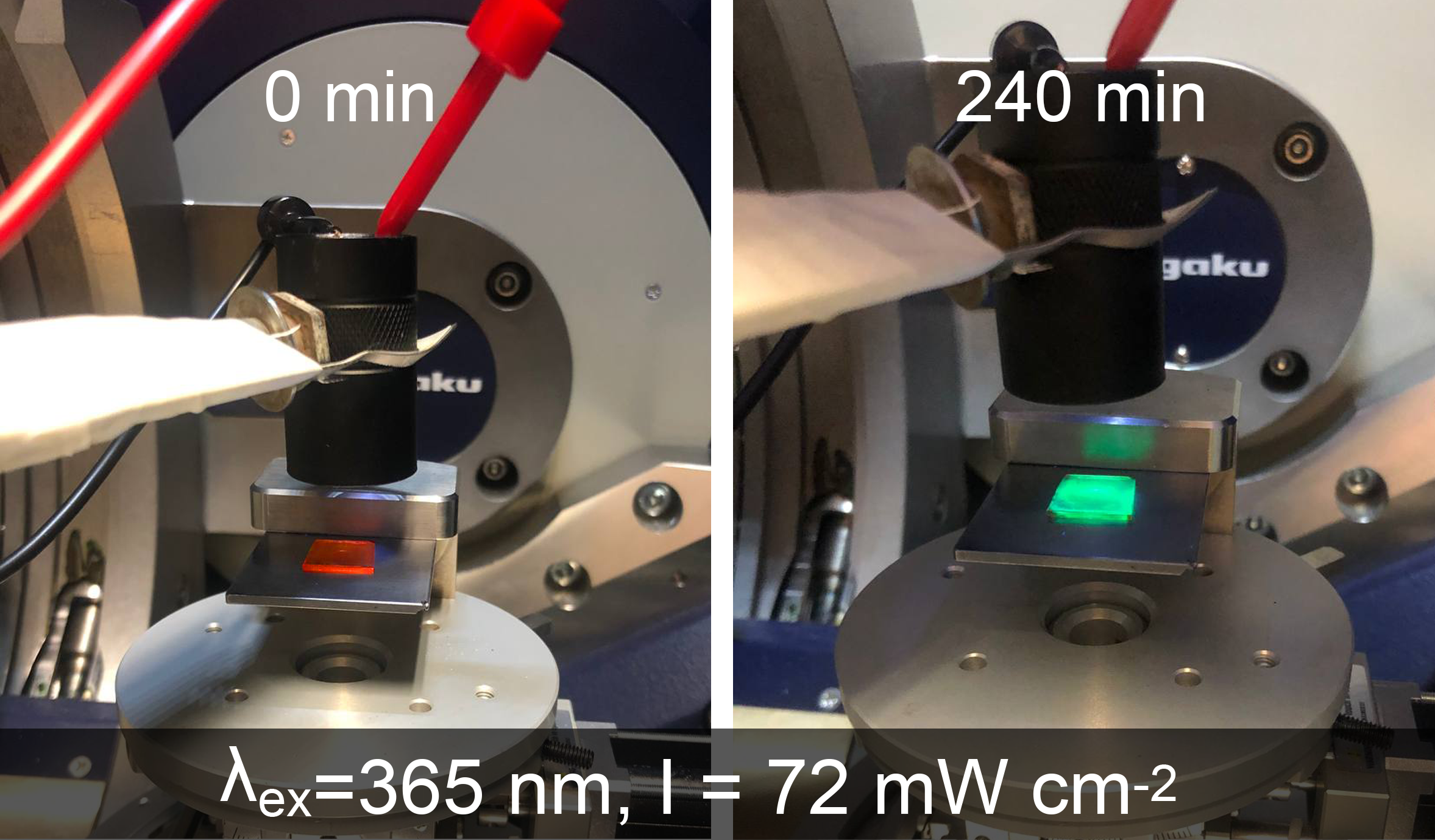}
\caption{Photographs of the $n=3$ quasi-2D perovskite film under continuous-wave UV illumination taken at the beginning and end of $in~situ$ XRD measurement.}
\label{}
\end{figure*}

\newpage
To estimate band gap values for BA$_2$MA$_{n-1}$Pb$_n$Br$_{n+1}$I$_{2n}$ ($n$ = 1--3) phases the deconvolution of their absorption spectra $\alpha(h\nu)$ is done (Fig. S2a,c,e). For an adequate approximation of excitonic absorption, we employ the following function describing a sum of discrete excitonic peaks:
\begin{equation}
{\alpha(\textit{h}\nu) \propto~ \frac{1}{\textit{h}\nu}} \Bigg[\sum_{j} \frac{2E_b}{j^3} {sech\Bigg(\frac{\textit{h}\nu-E_g+\frac{E_b}{j^2}}{\Gamma}\Bigg) \Bigg]}
\label{excitonic fit}
\end{equation}
where $E_g$ denotes the band gap, $E_b$ - exciton binding energy, $\Gamma$ stands for spectral broadening of exciton, and $j$ takes integer values from 1 to 3. A signal compensation before the band-to-band absorption cutoff is obtained when the fitting parameters are adjusted as follows:
\newline
i)	for $n$ = 1, $\Gamma$ = 0.11 eV, $E_b$ = 0.22 eV, $E_g$ = 2.98 eV;
\newline
ii)	for $n$ = 2, $\Gamma$ = 0.07 eV, $E_b$ = 0.17 eV, $E_g$ = 2.62 eV;
\newline
iii)	for $n$ = 3, $\Gamma$ = 0.06 eV, $E_b$ = 0.15 eV, $E_g$ = 2.45 eV.

The residual signal in the spectra describes band-to-band absorption and can be further used for obtaining corresponding Tauc plots (Fig. S2b,d,f). By extrapolating a linear region of [$\alpha h \nu$]$^2$($h \nu$) curves to the abscissa ([$\alpha h \nu$]$^2$=0) the same values of band gaps are derived.

\begin{figure*}
\includegraphics[width=1.\columnwidth]{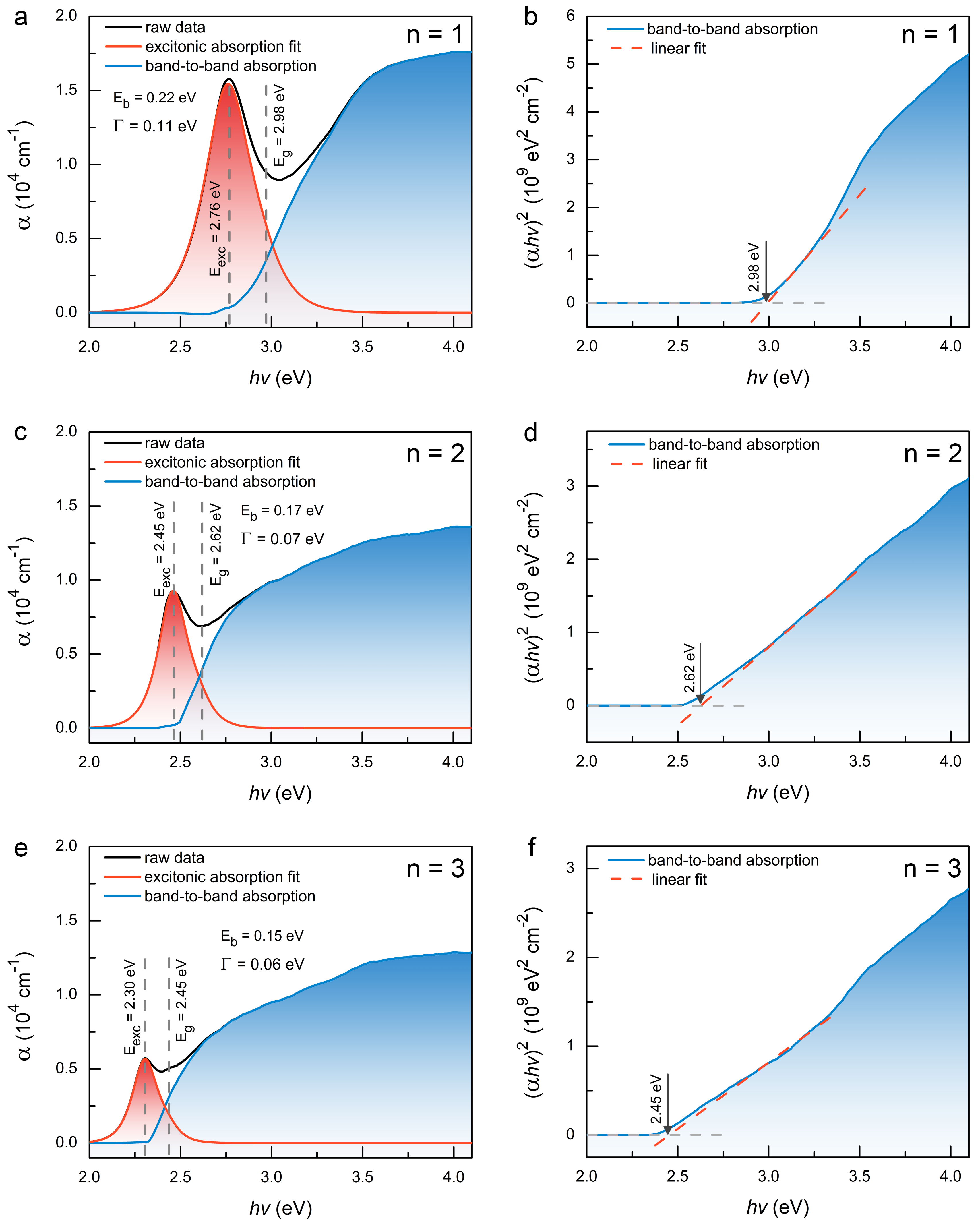}
\caption{(a,c,e) Deconvolution of absorption spectra for n = 1 (a), n = 2 (c), and n = 3 (e) thin films. (b,d,f) Tauc plots for n = 1 (b), n = 2 (d), and n = 3 (f) thin films.}
\label{}
\end{figure*}

\begin{figure*}
\includegraphics[width=1.\columnwidth]{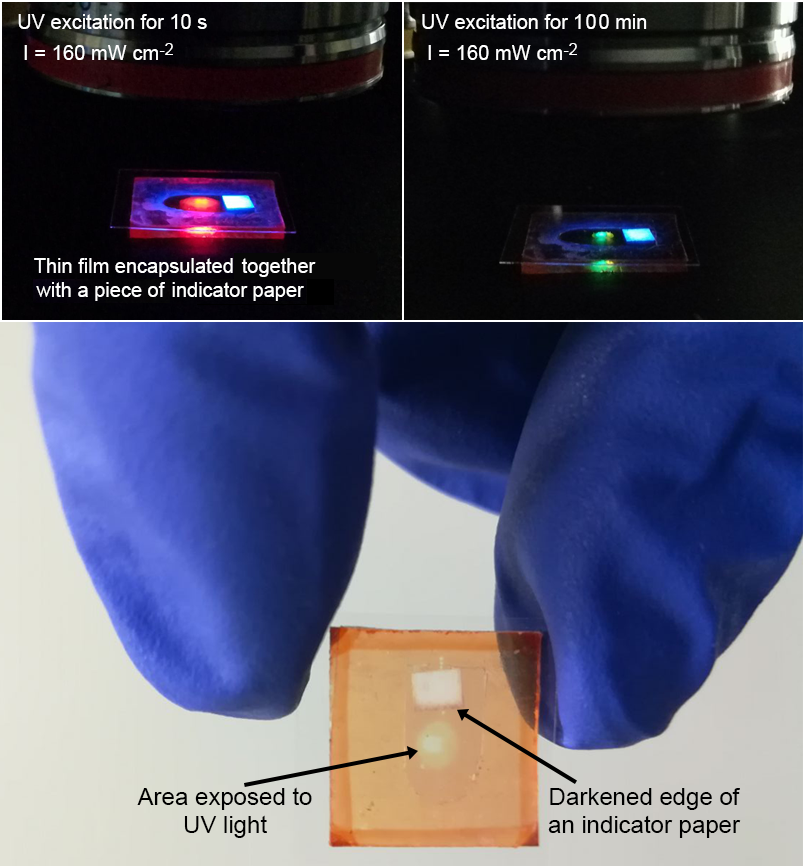}
\caption{Photographs of the $n=3$ quasi-2D perovskite film sealed together with a piece of potassium iodide indicator paper upon focused continuous-wave UV excitation with intensity of 160 mW cm$^{-2}$  for 10 s (left top image) and for 100 min (right top image). Bottom picture reveals a darkened edge of the indicator paper beside the exposed area of the film, proving I$_2$ species release during the illumination procedure.}
\label{}
\end{figure*}

\begin{figure*}
\includegraphics[width=1.\columnwidth]{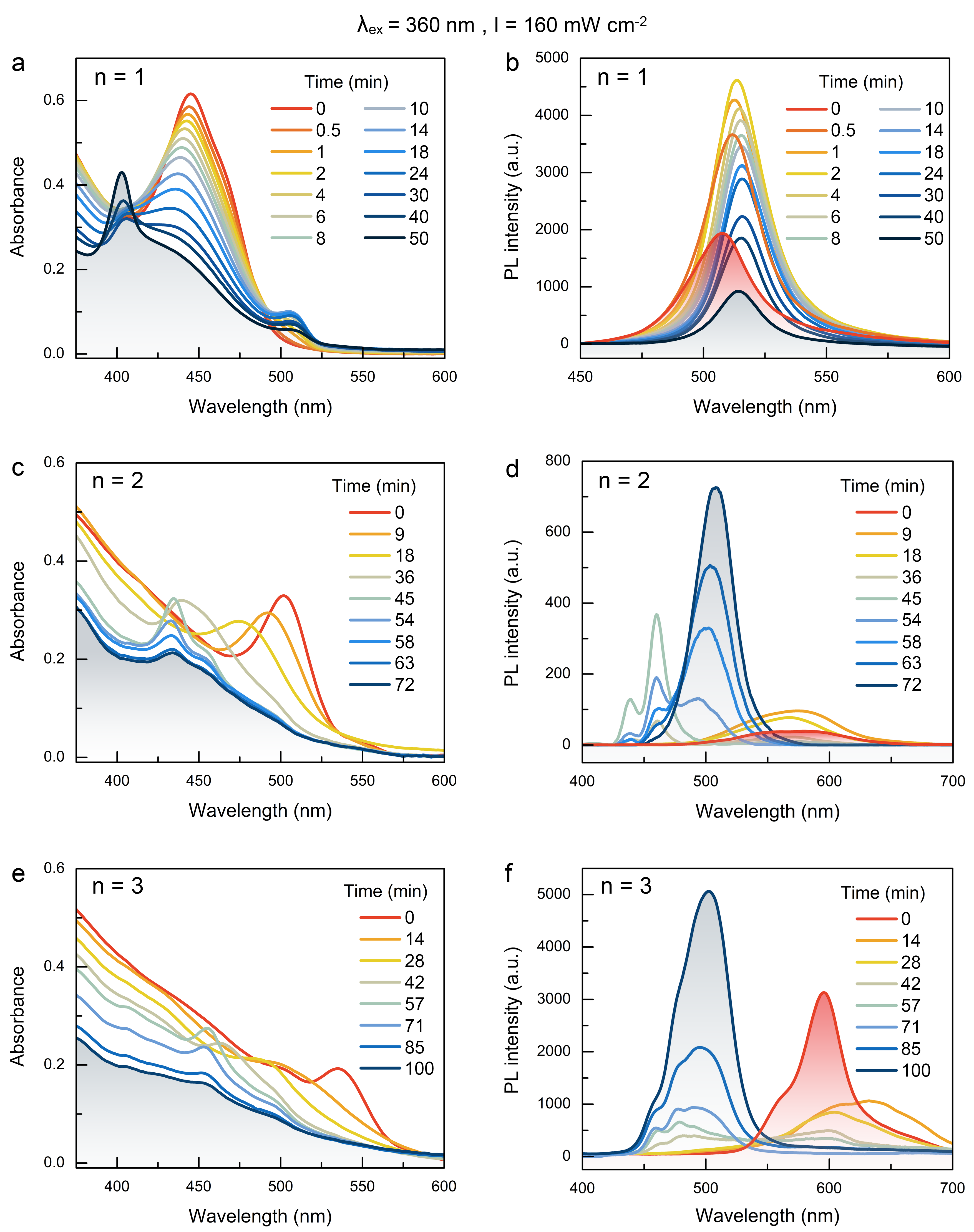}
\caption{Absorption and PL dynamics for $n$ = 1--3 films exposed to UV light.}
\label{}
\end{figure*}

\begin{figure*}
\includegraphics[width=1.\columnwidth]{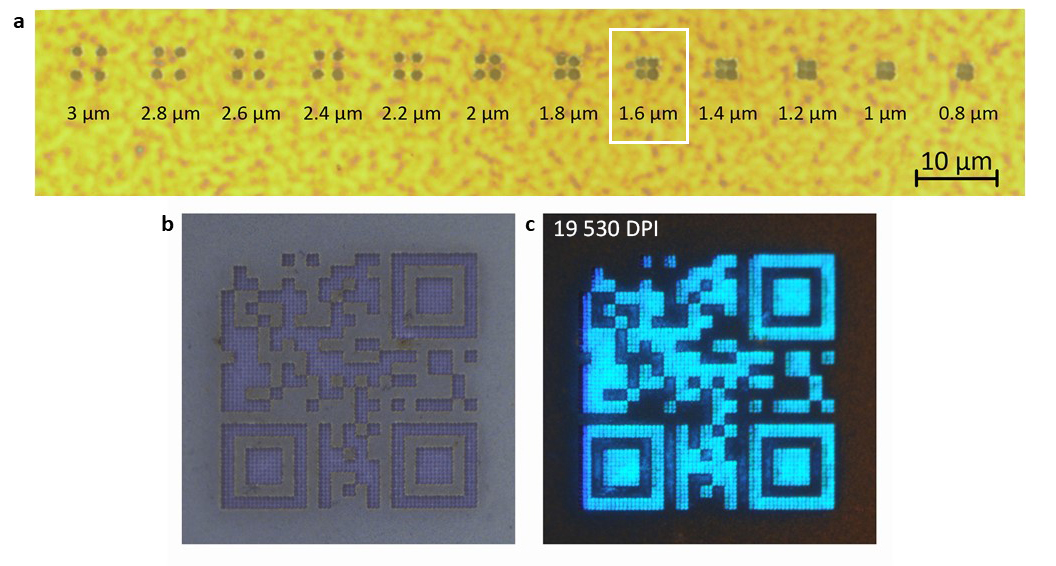}
\caption{(a) A bright-field image of a set of 2$\times$2 dots patterns on the $n=3$ quasi-2D perovskite film obtained on a direct laser writing setup. A dimension in $\mu$m below each pattern indicates the distance between the centers of 1 $\mu$m diameter dots. A white box defines the pattern demonstrating the best spatial resolution of 19,500 DPI. (b,c) High-resolution bright-field and fluorescent microimages of the QR code produced by using two-photon direct laser writing method.}
\label{}
\end{figure*}

\begin{figure*}
\includegraphics[width=1.\columnwidth]{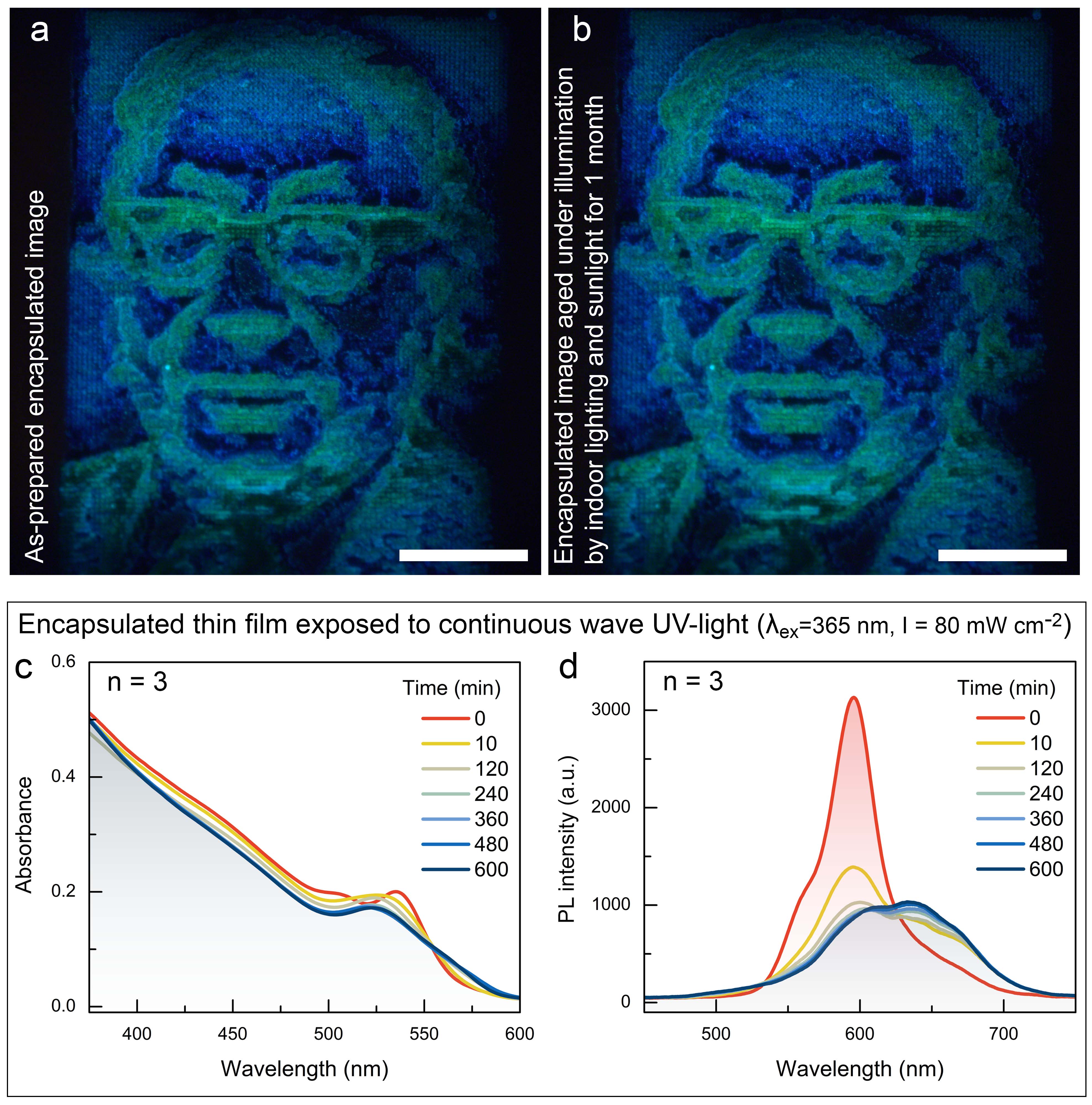}
\caption{(a,b) Pictures of an as-prepared (a) and aged (b) luminescent microimage encapsulated at ambient conditions (scale bar is 50 $\mu$m). (c,d) Absorption and PL spectral dynamics for $n$ = 3 film encapsulated at ambient conditions and then exposed to UV light for 600 min.}
\label{}
\end{figure*}

\begin{figure*}
\includegraphics[width=1.\columnwidth]{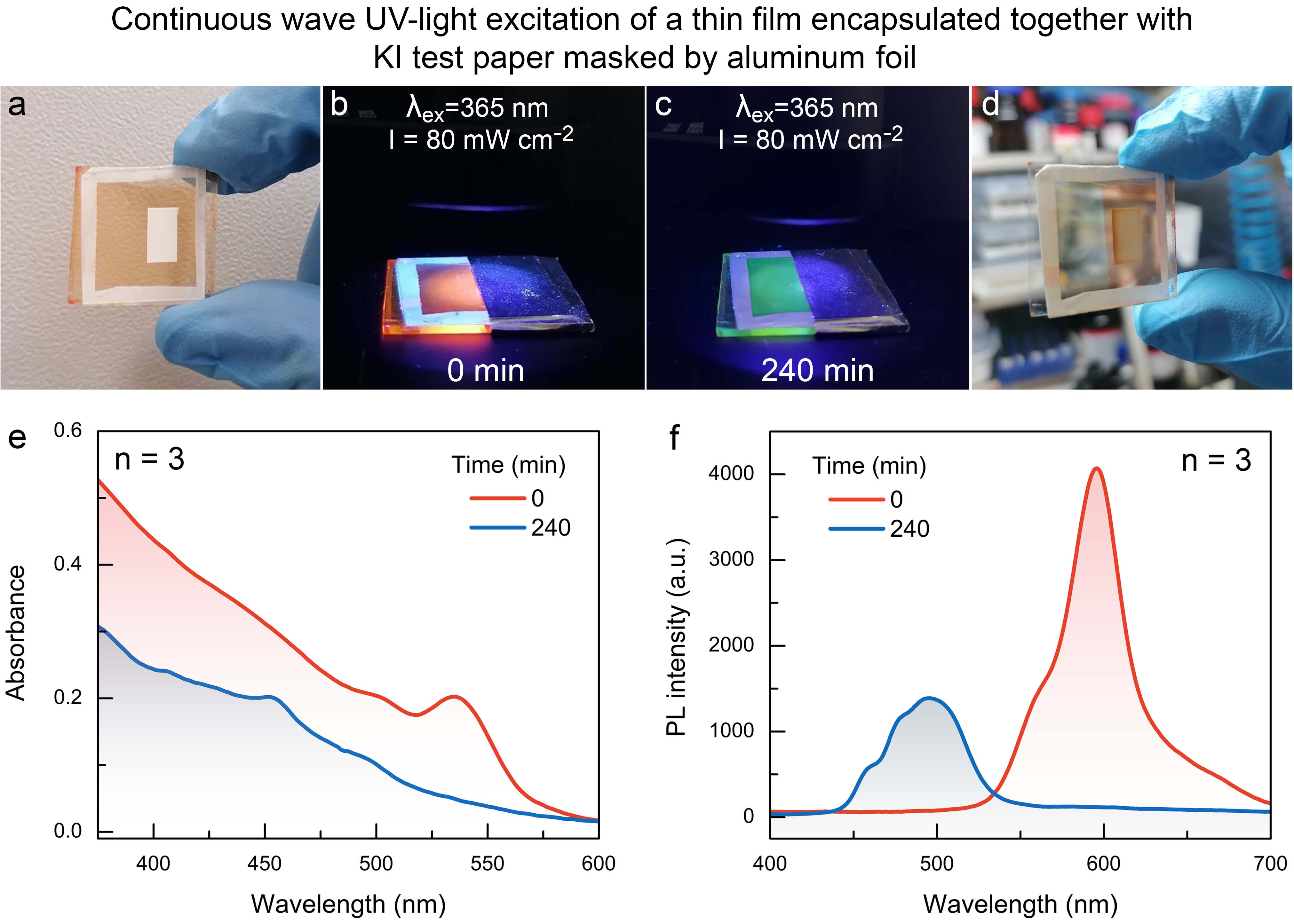}
\caption{(a) Picture of $n$ = 3 film encapsulated together with KI test paper in a N$_2$-filled glovebox. (b,c) Pictures showing PL color for a non-masked area of the film at the beginning of the UV-light exposure procedure (b) and after 240 min of continuous illumination (c). KI test paper is masked by aluminum foil. (d) Picture presenting a bleached part of the film along with a dark edge of the test paper as a result of the release of I$_2$ species from the former and their absorption by the latter. (e,f) The change in the absorption and PL spectra for the non-masked area of the film at the beginning of the UV-light exposure procedure and at its end.}
\label{}
\end{figure*}


